\documentclass[article,cleveref,autoref,thm-restate,anonymous]{article}
\usepackage[T1]{fontenc}
\usepackage[letterpaper]{geometry}
\usepackage{hyperref}
\usepackage{cite}
\usepackage{amsmath,amssymb,amsfonts}
\usepackage{cleveref}
\usepackage{graphicx}
\usepackage{textcomp}
\usepackage{xcolor}
\usepackage[linesnumbered,lined,ruled,noend]{algorithm2e}
\usepackage{subcaption}

\SetKwInput{KwRequire}{Prereq}
\SetKwInput{KwData}{Input}
\SetKwInput{KwResult}{Output}
\SetKwFor{PFor}{parfor }{do}{end}
\SetKwIF{If}{ElseIf}{Else}{if}{:}{elif}{else:}{}

\newcommand{\prefix}{\textsc{p}}
\newcommand{\clique}{\mathcal{C}}

\newcommand{\faces}{\mathcal{F}}

\newcommand{\myparagraph}[1]{ \noindent {\bf #1.}}

\definecolor{myblue}{RGB}{0,128,255}

\usepackage{colortbl}
\definecolor{Gray}{gray}{0.95}
\usepackage{xcolor}

\begin{document}

\title{Faster Parallel Triangular Maximally Filtered Graphs and Hierarchical Clustering}
\author{Steven Raphael \\MIT \and Julian Shun \\ MIT }
\date{}

\maketitle

\begin{abstract} \small\baselineskip=9pt Filtered graphs provide a powerful tool for data clustering. The triangular maximally filtered graph (TMFG) method, when combined with the directed bubble hierarchy tree (DBHT) method, defines a useful algorithm for hierarchical data clustering. This combined TMFG-DBHT algorithm has been shown to produce clusters with good accuracy for time series data, but the previous state-of-the-art parallel algorithm has limited parallelism.

This paper presents an improved parallel algorithm for TMFG-DBHT.
Our algorithm increases the amount of parallelism by aggregating the bulk of the work of TMFG construction together to reduce the overheads of parallelism. Furthermore, our TMFG algorithm updates information lazily, which reduces the overall work. 
We find further speedups by computing all-pairs shortest paths approximately instead of exactly in DBHT.
We show experimentally that our algorithm gives a 3.7--10.7x speedup over the previous state-of-the-art TMFG-DBHT implementation, while preserving clustering accuracy.\end{abstract}


\section{Introduction}\label{sec:intro}

Data clustering is a task that involves grouping data points by similarity. It has diverse applications across many fields \cite{clustering19}. 
Hierarchical clustering is a data clustering method that involves computing a \emph{dendrogram}, which is a tree of clusters with data points at the leaves. Hierarchical clustering is useful because this dendrogram provides additional information about the structure of the data, such as how different clusters are related to each other. 

\emph{Agglomerative clustering} algorithms construct the dendrogram by initializing one cluster per data point, and merging pairs of clusters until all clusters have been merged. Simple agglomerative clustering methods suffer from low accuracy~\cite{song2012}. 
To fix this problem, various methods have been proposed to create \emph{filtered graphs} from input datasets, where vertices correspond to data points, and edges correspond to correlations. Filtered graphs contain relatively few edges, in order to represent most important correlations in the dataset.
The triangular maximally filtered graph (TMFG)~\cite{Massara2017} is a type of filtered graph that is constructed by repeatedly connecting one uninserted vertex to all three vertices in a triangular face of the graph, where the vertex and face are chosen to maximize the sum of the newly inserted edges. The Directed Bubble Hierarchy Tree (DBHT)~\cite{song2012} method uses this graph to generate a dendrogram. It has been shown that the DBHT algorithm results in good clustering quality~\cite{musmeci2015relation, song2015multiscale}. 
More details on TMFG and DBHT are provided in Section~\ref{sec:background}.
Other methods for generating filtered graphs have been proposed (e.g., $k$-nearest neighbor graphs~\cite{Ruan2010} and minimum spanning trees~\cite{Mantegna1999,Tumminello2007}), but TMFG-DBHT has been shown to perform particularly well on time series data, which we focus on in this paper. 

Yu and Shun~\cite{yu2023parallel} design a parallel implementation of TMFG-DBHT that incorporated many optimizations over previous implementations (which were sequential). The main bottleneck in Yu and Shun's implementation of the TMFG and DBHT pipeline (TMFG-DBHT) is filtered graph construction. The original method of graph construction adds one vertex to the graph at a time, and there is a slow sorting step after each iteration. The more efficient implementation modifies the graph construction procedure by inserting several vertices at once, allowing the insertion steps (including the sorting steps) to run in parallel. This added parallelism increases the speed of graph construction on multi-core processors. However, there is a trade-off between speed and accuracy: increasing the number of vertices that are added per step lowers the quality of the filtered graph, which decreases the accuracy of the clustering algorithm. Yu and Shun's algorithm~\cite{yu2023parallel} sets the number of vertices added per step to 10 by default, which they found to give a good tradeoff between speed and clustering quality. In our experiments, we show that 
adding 10 vertices per step in their algorithm speeds up the overall runtime by a factor of 1.1--1.8, while only resulting in a slight decrease in accuracy on average, compared to the sequential algorithm.
This algorithm likely does not take full advantage of parallel processing: there is only a small amount of work that can be parallelized at a given time, causing the overhead of parallel processing to be relatively large. 
This paper further improves the speed of TMFG-DBHT by proposing various optimizations and algorithmic changes, the most significant of which involves changing the TMFG construction method to take more advantage of parallel processing. We discovered that the slowest part of each vertex insertion step is the sorting step that sorts a constant number of arrays for each newly added vertex. 
Unlike in the current state-of-the-art TMFG implementation, the new construction method no slow sorting steps that accompany each insertion. Instead, the insertion method is modified to allow all sorting to take place in a single sorting step at the beginning of TMFG construction. After this sorting step, the time it takes to insert a single vertex is so small that inserting several vertices at once can worsen performance due to additional overhead. Thus, the new construction method adds vertices serially.

The large amount of parallelism of our new algorithm reduces the overhead of parallel processing, greatly increasing the speed of graph construction. In addition, we show that our new construction method achieves comparable accuracy to~\cite{yu2023parallel}.
In addition, we made another change to speed up TMFG construction: using a heap instead of an array to store possible triangular faces that are part of the graph. The heap improves performance by allowing updates to these triangular faces to take place lazily, with the update to a given face only taking place when a face is at the root of the heap.

We made additional optimizations other than changing the TMFG construction method. The state-of-the-art TMFG-DBHT algorithm involves computing all-pairs shortest paths (APSP) on the constructed TMFG. APSP is not a bottleneck in their implementation, but since our new TMFG construction method makes TMFG construction faster, the speed of APSP is similar to that of TMFG construction in our implementation. As APSP now takes a significant portion of the total time, we propose an approximate version of APSP, which we found is significantly faster than the exact version, especially on large data sets. 
Finally, we use manual vectorization for certain parts of TMFG construction, along with a vectorized sorting algorithm from Google Highway, which give us a slight speedup. 

We experimentally evaluated our new implementations against the state-of-the-art TMFG-DBHT algorithm by Yu and Shun using a collection of time series and image data sets using a 48-core machine.
We show that our optimizations result in a 3.7--10.7 times speedup over the state-of-the-art algorithm, with 6\% better accuracy on average as measured by the Adjusted Rand Index scores. Our optimized algorithm achieves up to 34x self-relative speedup 
when running on a 48-core machine with two-way hyper-threading.
Our code is publicly available at \url{https://github.com/stevenraphael/par-filtered-graph-clustering/}.

\section{Background}\label{sec:background}

 \myparagraph{Definitions} 
A correlation (or similarity) matrix $S$ can be viewed as a complete graph, where $S[i,j]$ is the weight of edge $(i,j)$ in the graph.
A \emph{planar} graph is a graph that can be drawn on a plane with no crossing edges.

\myparagraph{PMFG, TMFG, and DBHT and Sequential Methods}
The Planar Maximally Filtered Graph (PMFG)~\cite{Tumminello2005} is a method to transform a correlation matrix of elements into a planar graph, where the vertices correspond to elements and the edges correspond to "important" correlations. The goal is to maximize the sum of the correlations (although doing so exactly is NP-Hard~\cite{giffin1984graph}). 

The Triangular Maximally Filtered Graph (TMFG)~\cite{Massara2017} is a faster method that constructs a planar graph, again where all faces are triangles, and the sum of all edges approximates that of the PMFG. 
Massara et al.~\cite{Massara2017} introduce a serial algorithm for constructing the TMFG, which works as follows.
Four vertices are selected as a starting 4-clique in the TMFG, resulting in 4 triangular faces. In every subsequent step, one vertex is added to the graph by connecting it to three vertices in an existing face. The vertex and face are chosen in a way that maximizes the sum of the three new edge weights. This sum is referred to as the "gain" of the face-vertex pair.

The Directed Bubble Hierarchy Tree~\cite{song2012,song2011nested} (DBHT) method performs a standard hierarchical clustering algorithm, such as single-linkage or complete-linkage, on the TMFG. The combination of these methods, TMFG-DBHT or PMFG-DBHT, has been shown to work well for clustering a variety of datasets~\cite{musmeci2015relation,wang2017multiscale, yen2021using, song2015multiscale,burton2015pathogenesis}. For more details, see~\cite{yu2023parallel}. 
The DBHT algorithm constructs a tree, called a bubble tree, by having a node per 4-clique in the TMFG. Every pair of 4-cliques that shares a triangular face is connected by a directed edge in the DBHT. Edge direction corresponds to which region--either inside or outside the triangular face--has stronger connections with the face. This DBHT is used to divide vertices into a hierarchy of groups. Group assignments are determined in multiple ways. The coarsest layer of groups involve "converging bubbles," nodes in the DBHT that have only incoming edges. Every vertex is assigned to a converging bubble. Each of these groups is subdivided into additional groups by assigning every vertex to a "bubble" (4-clique) based on connection strength. Connection strength in this case is determined by shortest-path distances in the TMFG. The The groups in each layer of the hierarchy are clustered using complete linkage, where distances are determined by the shortest paths in the TMFG. In order to determine distances for complete linkage, an all-pairs shortest paths algorithm is run on the TMFG.

\myparagraph{Parallel Methods}
Yu and Shun introduced an efficient implementation of TMFG construction~\cite{yu2023parallel} using parallelization (par-TMFG). 
par-TMFG adds multiple vertices to the TMFG during each step, where the vertices and corresponding faces are again chosen to maximize the gain of each face-vertex pair. This improves the running time of TMFG construction by increasing the amount of work that can be parallelized at each step. However, the parallelized version also results in a lower-quality graph: sometimes, sub-optimal face-vertex pairs are added to the TMFG that are not added in the serial TMFG algorithm. This is because parallelization reduces the number of faces that newly added vertices can be connected to. In particular, this algorithm creates a trade-off between speed and accuracy.

Yu and Shun additionally created an efficient implementation of DBHT, which involves some parallelization. The parallelized parts of DBHT include all-pair shortest path (APSP) computation and complete-linkage clustering. APSP is computed using many instances of Dijkstra's algorithm, one instance per starting vertex. Each instance of Dijkstra's algorithm is run in parallel. Complete-linkage clustering is parallelized using the algorithm created by Yu et al.~\cite{Yu2021}.




In the TMFG construction algorithms that follow, a face-vertex pair refers to a pair of objects: a face in the incomplete TMFG, and an uninserted vertex that can be potentially be connected to that face. At each step of the TMFG construction algorithms, a face-vertex pair is chosen to be inserted into the TMFG, with the uninserted vertex being connected to all three vertices in the face.
\section{Related Work}

Hierarchical agglomerative clustering (HAC) algorithms are widely used to perform hierarchical clustering. The work most related to ours is the parallel TMFG-DBHT algorithm by Yu and Shun~\cite{yu2023parallel}, which we have discussed earlier.
Besides DBHT combined with TMFG~\cite{yu2023parallel,Massara2017,song2011nested,song2012}, many other parallel HAC algorithms have been proposed (e.g.,~\cite{mullner2013fastcluster, Yu2021, wang2021fast, terahac, rajasekaran2005efficient}).
DBHT combined with TMFG has been shown to have better accuracy than single-linkage and average-linkage HAC~\cite{song2012,musmeci2015relation}. 
Standard HAC methods can be sensitive to small changes in the dataset, and there have methods proposed to address this issue~\cite{Gagolewski_2016, balcan2014robust}.

In addition to HAC, other parallel hierarchical clustering methods, such as partitioning methods (e.g.,~\cite{dash2004efficient,  mao2015parallel}) or density-based methods (e.g.,~\cite{Campello2015, wang2021fast}) have been designed. 
Wang et al.~\cite{wang2021fast} design parallel algorithms for HDBSCAN$^*$, a hierarchical version of DBSCAN~\cite{Ester1996}.
Cohen-Addad et al.~\cite{cohen2019hierarchical} propose a parallel algorithm for the hierarchical $k$-median problem in Euclidean space. 
Dash et al.~\cite{dash2004efficient} propose a parallel algorithm that partitions data into overlapping cells and then performs clustering on the cells.
Mao et al.~\cite{mao2015parallel} and Wang et al.~\cite{wang2011parallel} present hierarchical clustering algorithms for specific data types.
The techniques in these papers are generally different than ours, since the clustering approaches are different.

PMFG and TMFG give approximate solutions to the weighted maximal planar graph (WMPG) problem, which has been shown to be NP-complete~\cite{giffin1984graph}.
Other approximate algorithms and heuristics for WMPG have also been proposed~\cite{kataki2020new,osman2003greedy,dyer1985analysis,Cimikowski1995,eades1982efficient,Calinescu2003}.


\section{Our New Algorithms}
This section describes our two new algorithms, which are improvements to Yu and Shun's parallel TMFG algorithm~\cite{yu2023parallel}: correlation-based TMFG and heap-based TMFG. Here we refer to Yu and Shun's TMFG construction algorithm as $\textsc{orig-tmfg}$.

\subsection{Correlation-based TMFG}
\label{sec:corr_tmfg}
We define a new TMFG construction algorithm, which we refer to as $\textsc{corr-tmfg}$. 
First note that, at every step, $\textsc{orig-tmfg}$ creates a face-vertex pair for every face by picking uninserted vertices, picking the vertex that would maximize the edge sum when connected to a given face, and then selects one or more face-vertex pairs to insert in the graph based on which insertions would maximize the edge sum. $\textsc{corr-tmfg}$ differs from $\textsc{orig-tmfg}$ in the way that the face-vertex pairs are chosen. First, $\textsc{corr-tmfg}$ pre-computes, for every vertex, a sorted list of correlations to every other vertex. Instead of directly computing the maximum gain from a face to any vertex, $\textsc{corr-tmfg}$ instead uses the sorted lists of correlations to compute the maximum correlation from each of the face's three vertices to any other vertex. This results in a list of up to three candidate vertices for the face-vertex pair. The algorithm then computes the vertex with the maximum gain out of the candidates. 

$\textsc{corr-tmfg}$ is intended to boost the speed of graph construction. It does this by replacing the sorting steps that are interspersed in every step of $\textsc{orig-tmfg}$ with one sorting step at the beginning of the algorithm, with many arrays that can be sorted in parallel. Thus, the correlation-based algorithm can perform more work in parallel, leading to less computational overhead.

\begin{figure}[!t]
    \centering
    \includegraphics[width = \columnwidth]{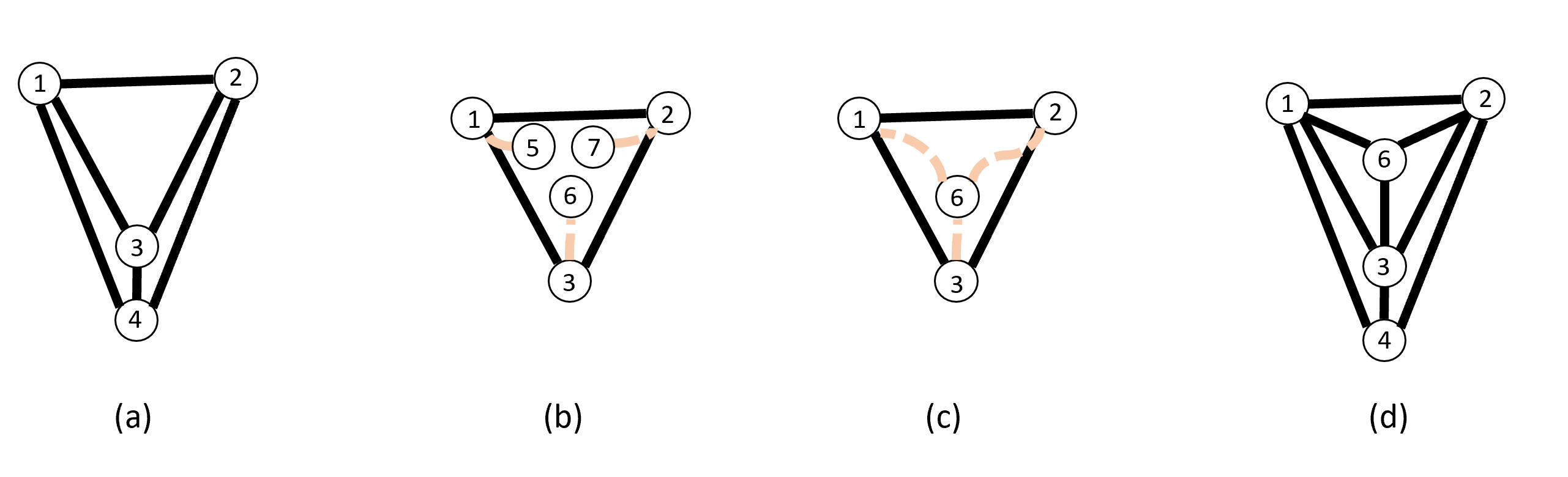}
    \caption{This figure shows an example of the first iteration of the \textsc{corr-tmfg} algorithm. (a) The initial 4-clique in the TMFG. (b) For each face, the closest vertex to each of the face's three vertices is found (the subfigure shows this process for the face $\{1,2,3\}$). (c) Among the three closest vertices, the vertex with maximum gain is selected for that face to form a face-vertex pair (for face $\{1,2,3\}$, this is vertex $6$). (d) Among all face-vertex pairs, the one with maximum gain is added to the TMFG (this is vertex $6$). }
    \label{fig:algo-example}
\end{figure}

Like $\textsc{orig-tmfg}$, $\textsc{corr-tmfg}$ can insert multiple vertices at once. We denote the number of vertices added together as the \emph{prefix size}. In practice, however, we find that we get the most performance benefit by only inserting one vertex per round. 
An example iteration of our algorithm is shown in \cref{fig:algo-example}.
In the rest of the subsection, we provide a detailed overview of $\textsc{corr-tmfg}$. 

\SetKwIF{If}{ElseIf}{Else}{if~(\endgraf}{\endgraf)~then}{else if}{else}{end if}%

\begin{algorithm}[!t]
\DontPrintSemicolon
\fontsize{7pt}{7pt}\selectfont
\KwData{An $n \times n$ similarity matrix $S$ and a prefix size \prefix \ $\geq 1$}

\tcp*[h]{The four vertices with the largest total sum across its row in $S$}\;
$\clique = \{v_1, v_2, v_3,  v_4\}$\;  \label{alg:partmfg:first4}
$\mathcal{E} = \{(v_1, v_2)
(v_1, v_3),
(v_1, v_4),
(v_2, v_3),
(v_2, v_4),
(v_3, v_4)\}$\;
\tcp*[h]{The four initial faces}\;
$\faces  =\{\{v_1,v_2,v_3\},\{v_1,v_2,v_4\},\{v_1,v_3,v_4\},\{v_2,v_3,v_4\}\}$\; 
$V_{all} = \{v_1, \dots, v_n\}$\; \label{alg:partmfg:allv}
$V_{rem} = \{v_5, \dots, v_n\}$\; \label{alg:partmfg:restv}
\tcp*[h]{Sort correlations by similarity}\;

\PFor{$v\in V_{all}$}{\label{alg:partmfg:sortloop}
    Sort $[(S[v,u], u)$ for $u\in V_{all}]$ by similarity\label{alg:partmfg:sort}
}
\tcp*[h]{Compute vertex with highest correlation for each vertex in each of the 4 initial faces}\;
$\textsc{MaxCorrs} = [\operatorname*{argmax}_{u \in V_{rem}}(S[v,u])$ for $v\in V_{all}]$\;\label{alg:partmfg:maxcorrs}
\PFor{$f\in \faces$}{\label{alg:partmfg:vmfor}
    $V_{m}[f] = [\textsc{MaxCorrs}[v]$ for $v\in f]$\;\label{alg:partmfg:vm}
}

\tcp*[h]{Compute best vertex for each of the 4 initial faces and construct \textsc{Gains} array}\;
$\textsc{Gains} = [\operatorname*{argmax}_{u \in V_{m}[f]}(\sum_{v \in f} S[v,u])$ for $f\in \faces ]$ \; \label{alg:partmfg:initgain}

\While{$|V_{rem}|>0$}{\label{alg:partmfg:while}
    $L$ = the $\prefix$ vertex-face pairs with the largest gains in $\textsc{Gains}$.\; \label{alg:partmfg:getprefix}
    For vertices paired with multiple faces in $L$, only keep the pair with maximum gain.\; \label{alg:partmfg:conflict}
    $V_{rem} = V_{rem} \setminus \{\text{vertices} \in L\}$\; \label{alg:partmfg:restv2}
    \PFor{$(v,t= \{v_x, v_y, v_z\}) \in L$ }{\label{alg:partmfg:for}
        $\mathcal{E} =\mathcal{E} \cup \{(v, v_x), (v, v_y), (v, v_z)\}$\;\label{alg:partmfg:13}
        $\faces = \faces \cup \{\{v, v_x, v_y\}, \{v, v_y, v_z\}, \{v, v_x, v_z\}\} \setminus t $\;\label{alg:partmfg:14}
        $\faces_{update} = t \in$ $\{t': \textsc{Gains} [t']=v\} \cup  \{\{v, v_x, v_y\}, \{v, v_y, v_z\}, \{v, v_x, v_z\}\}$\; \label{alg:partmfg:getfaces}
        $V_{update} = $ $\bigcup[[v\in t]$ for $t \in$ $\faces_{update}]$\; \label{alg:partmfg:getvertices}
        \tcp*[h]{ Compute un-inserted vertex with highest correlation to each vertex that needs to be updated}\;
        \PFor{$v \in V_{update}$ }{ \label{alg:partmfg:updatecorr}
            $\textsc{MaxCorrs}[v] = \operatorname*{argmax}_{u \in V_{rem}}(S[v,u])$\;
            \label{alg:partmfg:updatecorrloop}
        }

\tcp*[h]{Compute best vertex for each face that needs to be updated and add pairs to \textsc{Gains} array}\;
        \PFor{$f \in$ $\faces_{update}$ }{ \label{alg:partmfg:updategain}
            $V_{m}[f] = [\textsc{MaxCorrs}[v]$ for $v\in f]$\;\label{alg:partmfg:updatevm}
            $\textsc{Gains}[f] = \operatorname*{argmax}_{u \in V_{m}[f]}(\sum_{v \in t} S[v,u])$\;
            \label{alg:partmfg:updategainloop}
        }
    }
}
\Return $\mathcal{E}$\; \label{alg:partmfg:return}
\caption{Correlation-based TMFG}\label{alg:partmfg}
\end{algorithm}

The pseudocode for $\textsc{corr-tmfg}$ is shown in algorithm \ref{alg:partmfg}.
Lines~\ref{alg:partmfg:first4}--\ref{alg:partmfg:restv} initialize the TMFG with the four best vertices. $\clique$ is the first clique of vertices to be inserted into the DBHT, $\mathcal{E}$ contains the edges in the TMFG, $\faces$ contains the triangular faces in the TMFG, $V_{all}$ contains all vertices regardless of whether they have been inserted, and $V_{rem}$ contains the uninserted vertices.
The loop on Lines~\ref{alg:partmfg:sortloop}--\ref{alg:partmfg:sort} creates, for every vertex $v$, an array that sorts all other vertices by similarity with $v$. Next, Line~\ref{alg:partmfg:maxcorrs} uses this array to create $\textsc{MaxCorrs}$, which contains, for every vertex $V$, the uninserted vertex with the highest similarity to $v$. The loop on Lines~\ref{alg:partmfg:vmfor}--\ref{alg:partmfg:vm} uses $\textsc{MaxCorrs}$ to define an array $V_m$ that contains, for each face $f$ in the TMFG, a set of up to three candidate vertices associated with $f$. Specifically, $V_m[f]$ contains the uninserted vertices with highest similarity to some vertex in $f$. Line~\ref{alg:partmfg:initgain} creates the $\textsc{Gains}$ array by computing, for every face $f$, the candidate vertex in $V_m[f]$ with the highest gain (defined as the sum of similarities to each vertex in $f$).\\

Next, the while loop on Line~\ref{alg:partmfg:while} contains the main part of the algorithm. It inserts additional vertices into the TMFG while there are still vertices to be inserted. On Line~\ref{alg:partmfg:getprefix}, the algorithm finds the vertices in $\textsc{Gains}$ with the maximum gain using parallel sorting and stores these vertices in $L$. Our most efficient version of the algorithm adds vertices one at a time, but it is possible to add multiple vertices in a single iteration. For vertices that are paired with multiple faces, we only keep the pair with the maximum gain, which can be computed using a parallel filter and parallel sorting on Line~\ref{alg:partmfg:conflict}.
The vertices are removed from $V_{rem}$ on Line~\ref{alg:partmfg:restv2}. Every time a new vertex is added, the algorithm needs to update $\textsc{Gains}$, which is computed from $V_m$. It loops through all added face-vertex pairs in the Loop on line~\ref{alg:partmfg:for}. For each added pair, it adds the new edges to the graph on Line~\ref{alg:partmfg:13} and new faces to $\faces$ on Line~\ref{alg:partmfg:14}. On Line~\ref{alg:partmfg:getfaces}, it calculates $\faces_{update}$, the list of faces $f$ where $V_m[f]$ needs to be computed or updated after the new vertices are inserted. From there, on Line~\ref{alg:partmfg:getvertices}, the algorithm calculates $V_{update}$, the list of vertices $v$ contained in faces in $\faces_{update}$, such that $\textsc{MaxCorrs}[v]$ needs to be calculated for updating $V_m$. For each of these vertices, the algorithm recomputes each $\textsc{MaxCorrs}[v]$ on Lines~\ref{alg:partmfg:updatecorr}--\ref{alg:partmfg:updatecorrloop}. On Line~\ref{alg:partmfg:updategain}, in parallel for each face $f$ to be updated, the algorithm recalculates $V_m[f]$ on Line~\ref{alg:partmfg:updatevm}, and finally uses this to update $\textsc{Gains}[f]$ on Line~\ref{alg:partmfg:updategainloop}. This completes one iteration of TMFG vertex insertion.
If only one vertex is inserted per round, the update step can be somewhat simplified: $V_{update}$ can simply bet set to contain the four vertices in the newly added face-vertex pair.
Note that in our implementation, when a vertex is added to the TMFG, we also add it to the bubble tree that is used in DBHT (this is not shown in the pseudocode).

\subsection{Heap-based TMFG}\label{heap_tmfg}
This subsection describes a second modification to the TMFG algorithm, which we call heap-based TMFG, or $\textsc{heap-tmfg}$. Using the correlation-based method of computing the best vertex for a given face, the algorithm can be further improved in performance by lazily updating the best face-vertex pairs using a heap structure. This new method works by storing the best face-vertex pair for each face in a max-heap sorted by gain. Unlike $\textsc{corr-tmfg}$, the face-vertex pairs are not updated whenever a vertex is added to the growing TMFG; a face-vertex pair is only updated when it gets popped from the max-heap and the vertex has already been added to the TMFG. In this case, the updated face-vertex pair is re-inserted into the heap. The algorithm might repeatedly pop pairs from the heap until it finds an updated face-vertex pair with an uninserted vertex, which is used as the next vertex added to the TMFG. 

Using a heap increases the performance by reducing the number of times the best face-vertex pairs are updated. Furthermore, the quality of the resulting graphs using the heap-based method is only slightly different from the graphs produced without using a heap. We confirmed this by comparing the edge sums of the resulting graphs, as described in the experiments section. Intuitively, the best face-vertex pair selected by the heap in each round should usually be the same as the best pair selected without the heap-based method. The only exception is when there is some node below the root of the heap that would have a higher gain than the root node if it were updated. This situation requires that the gain of a face node increases after that node is updated. If updating a node always selects the uninserted vertex with the highest gain, the gain should never increase after an update, since a strictly worse vertex is always chosen. Assuming the correlation-based method usually selects the vertex with the highest gain, it should be rare for a face node to have a higher gain after an update, and thus the above exception is rare.

\begin{algorithm}[!t]
\DontPrintSemicolon
\fontsize{7pt}{7pt}\selectfont
\KwData{An $n \times n$ similarity matrix $S$ and a prefix size \prefix \ $\geq 1$}
\tcp*[h]{The four vertices with the largest total sum across its row in $S$}\;
$\clique = \{v_1, v_2, v_3,  v_4\}$\;  \label{alg:htmfg:first4}
$\mathcal{E} = \{(v_1, v_2)
(v_1, v_3),
(v_1, v_4),
(v_2, v_3),
(v_2, v_4),
(v_3, v_4)\}$\;
\tcp*[h]{The four initial faces}\;
$\faces  =\{\{v_1,v_2,v_3\},\{v_1,v_2,v_4\},\{v_1,v_3,v_4\},\{v_2,v_3,v_4\}\}$\; 
$V_{all} = \{v_1, \dots, v_n\}$\; \label{alg:htmfg:allv}
$V_{rem} = \{v_5, \dots, v_n\}$\; \label{alg:htmfg:restv}
\tcp*[h]{Sort correlations by similarity}\;

\PFor{$v\in V_{all}$}{
    Sort $[(S[v,u], u)$ for $u\in V_{all}]$ by similarity
}
$\textsc{MaxCorrs} = [\operatorname*{argmax}_{u \in V_{rem}}(S[v,u])$ for $v\in V_{all}]$\;
\tcp*[h]{Compute vertex with highest correlation for each vertex in each of the 4 initial faces}\;
\PFor{$f\in \faces$}{
    $V_{m}[f] = [\textsc{MaxCorrs}[v]$ for $v\in f]$\;
    \label{alg:htmfg:corrsinit}
}
\tcp*[h]{Put face-vertex pairs in a heap}\;
$\textsc{Gains} = [\operatorname*{argmax}_{u \in V_{m}[f]}(\sum_{v \in f} S[v,u])$ for $f\in \faces ]$ \; \label{alg:htmfg:initgain}
Convert $\textsc{Gains}$ into max-heap sorted by $(\sum_{u \in t} S[u,v])$ for a given face-vertex pair $(v,t)$\;  \label{alg:htmfg:convertgain}

\While{$|V_{rem}|>0$}{\label{alg:htmfg:while}
    $(v,t= \{v_x, v_y, v_z\}) = \text{pop}(\textsc{Gains})$\; \label{alg:htmfg:pop}

    \If{$v\in V_{rem}$}{\label{alg:htmfg:if}
        $V_{rem} = V_{rem} \setminus v$\; \label{alg:htmfg:restv2}
        $\mathcal{E} =\mathcal{E} \cup \{(v, v_x), (v, v_y), (v, v_z)\}$\;\label{alg:htmfg:13}
        $\faces = \faces \cup \{\{v, v_x, v_y\}, \{v, v_y, v_z\}, \{v, v_x, v_z\}\} \setminus t $\;\label{alg:htmfg:14}
        \tcp*[h]{Compute un-inserted vertex with highest correlation to each vertex in the face}\;
        \PFor{$v \in \{v, v_x, v_y, v_z\}$ }{\label{alg:htmfg:updatecorrloop-start}
            $\textsc{MaxCorrs}[v] = \operatorname*{argmax}_{u \in V_{rem}}(S[v,u])$\;
            \label{alg:htmfg:updatecorrloop}
        }
        \tcp*[h]{Compute best vertex for each face and add pairs to heap}\;
        \PFor{$f \in$ $\{\{v, v_x, v_y\}, \{v, v_y, v_z\}, \{v, v_x, v_z\}\}$ }{             \label{alg:htmfg:updategainloopstart}
            $V_{m}[f] = [\textsc{MaxCorrs}[v]$ for $v\in f]$\;
            $\text{push}(\textsc{Gains}, (\operatorname*{argmax}_{u \in V_{m}[f]}(\sum_{v \in f} S[v,u]), f))$\;  \label{alg:htmfg:updategainloop}
        }
    }
    \Else{\label{alg:htmfg:else}
    \tcp*[h]{Compute un-inserted vertex with highest correlation to each vertex in the face}\;
        \PFor{$v \in \{v_x, v_y, v_z\}$ }{ \label{alg:htmfg:updatecorr2}
            $\textsc{MaxCorrs}[v] = \operatorname*{argmax}_{u \in V_{rem}}(S[v,u])$\;
            \label{alg:htmfg:updatecorrloop2}
        }
        $V_{m}[t] = [\textsc{MaxCorrs}[v]$ for $v\in t]$\;
        $v'=\operatorname*{argmax}_{u \in V_{m}[t]}(\sum_{v \in t} S[v,u])$\;\label{alg:htmfg:gaincalc}
        $\text{push}(\textsc{Gains}, (v', t))$\; \label{alg:htmfg:insert}
        \label{alg:htmfg:updategainloop2}
    }
}
\Return $\mathcal{E}$\; \label{alg:htmfg:return}
\caption{Heap-based TMFG}\label{alg:htmfg}
\end{algorithm}

We now describe the details of $\textsc{heap-tmfg}$, whose pseudocode is shown in algorithm \ref{alg:htmfg}. This method can only insert one vertex at a time, unlike the previous algorithms.
Lines~\ref{alg:htmfg:first4}--\ref{alg:htmfg:corrsinit} of $\textsc{heap-tmfg}$ are the same as $\textsc{corr-tmfg}$. 
$\textsc{Gains}$ is a max-heap, initialized on Lines~\ref{alg:htmfg:initgain}--\ref{alg:htmfg:convertgain}, which is sorted by the gain of each face-vertex pair in the heap. The algorithm inserts exactly one vertex at a time, by taking the maximum face-vertex pair $(v,t)$ in the $\textsc{Gains}$ heap (Line~\ref{alg:htmfg:pop}), until there are no remaining vertices to insert. If this pair contains a vertex that has already been inserted into the TMFG, we reach the else-clause on Line~\ref{alg:htmfg:else} and the best vertex for $t$ in the $\textsc{Gains}$ heap is re-computed on Lines~\ref{alg:htmfg:updatecorr2}--\ref{alg:htmfg:gaincalc} in a similarly way to \textsc{corr-tmfg} and the updated pair is re-inserted into the $\textsc{Gains}$ heap on Line~\ref{alg:htmfg:insert}. Otherwise, the if-statement on Line~\ref{alg:htmfg:if} is satisfied and the face-vertex pair is inserted into the TMFG.

When the new vertex is inserted, the best vertex for each face is not updated except for the three new faces created by inserting the vertex. The new edges are declared on Line~\ref{alg:htmfg:13}, and the faces are declared on Line~\ref{alg:htmfg:14}. The $\textsc{MaxCorrs}$ array is only updated for the four vertices in the face-vertex pair; this is done in a loop on Lines~\ref{alg:htmfg:updatecorrloop-start}--\ref{alg:htmfg:updatecorrloop}. In the loop on Lines~\ref{alg:htmfg:updategainloopstart}--\ref{alg:htmfg:updategainloop}, the face-vertex pairs corresponding to the newly inserted faces are fully computed. These face-vertex pairs are then inserted into the heap.

\subsection{Other Optimizations}\label{sec:opt}

We implemented a few other optimizations to the TMFG-DBHT procedure.
First, one issue is that, in order to update the maximum correlation for a given vertex when its current best vertex is added to the TMFG, the algorithm needs to iterate across the list of vertices until it reaches an uninserted vertex. To speed up this iteration, we implemented manual vectorization for AVX2 and AVX512.\\
We also improved the performance of the all-pairs shortest path (APSP) stage of DBHT-TMFG. We did this by switching to an approximate hub-based APSP algorithm. This algorithm only computes all shortest paths to and from a limited subset of vertices. To compute the distance between an arbitrary pair of vertices, the algorithm computes the distance from the start vertex to any hub. It then finds all points within a certain radius that depends on the distance to the hub. If the ending point is too far away, the algorithm approximates the shortest path distance by adding the hub-starting vertex distance to the hub-ending vertex distance. The exact parameters for the approximate APSP algorithm were selected arbitrarily, but we got a significant speed boost without sacrificing accuracy.

\section{Experiments}\label{sec:exp}
\myparagraph{Testing Environment} We perform experiments on
a \texttt{c5.24xlarge} machine on Amazon EC2, with 48
hyper-threaded Intel Xeon Platinum 8275CL (3.00GHz) CPU cores, or 96 vCPUs. It has 192 GiB of memory. By default, we use all cores with hyper-threading.  Our code uses ParlayLib~\cite{blelloch2020parlaylib} 
for parallelism and is compiled using the \texttt{g++} compiler (version 11.4.0)
with the \texttt{-O3} flag.

\myparagraph{Implementations}
\textsc{par-tdbht-1} is the previous implementation of parallel TMFG and DBHT by Yu and Shun with a prefix size of 1 for TMFG construction~\cite{yu2023parallel}. \textsc{par-tdbht-10} is a version of this algorithm with a prefix size of 10, and \textsc{par-tdbht-200} is a version with prefix size 200. \textsc{corr-tdbht} consists of \textsc{par-tdbht}, with TMFG construction replaced by an implementation of $\textsc{corr-tmfg}$. The prefix size is 1 for this algorithm. \textsc{heap-tdbht} also consists of \textsc{par-tdbht}, with TMFG construction replaced by an implementation of $\textsc{heap-tmfg}$. Finally, \textsc{opt-tdbht} is uses our heap-based TMFG algorithm with all of the optimizations introduced in~\Cref{sec:opt}.

\myparagraph{Evaluation}
We evaluate the clustering quality using the Adjusted Rand Index (ARI)~\cite{Hubert1985} score. 
Let $n_{ij}$ be the number of objects in the ground truth cluster $i$ and the cluster $j$ that is generated by the algorithm. Let $n_{i*}=\sum_j n_{ij}$, $n_{*j}=\sum_i n_{ij}$, and $n=\sum_i n_{i*}$. The ARI score is defined as 
$\frac{\sum_{i,j} {n_{ij}\choose 2} - [\sum_i {n_{i*}\choose 2} \sum_j {n_{*j}\choose 2} ] / {n\choose 2}}{\frac{1}{2}[\sum_i {n_{i*}\choose 2} +\sum_j {n_{*j}\choose 2} ] - [\sum_i {n_{i*}\choose 2} \sum_j {n_{*j}\choose 2} ]  /   {n \choose 2}}$.
The maximum ARI score
is $1$, corresponding to a perfect match. For random assignments, the expected ARI score is $0$.
When computing the ARI for the hierarchical methods, we cut the
dendrogram such that the number of resulting clusters is the same as the
number of ground truth clusters.

\begin{table}[!t]
\centering
\footnotesize
\begin{tabular}{ccccc}
ID & Name & $n$ & $L$ & Number of classes \\ \hline 
1 & CBF & 930 & 128 & 3 \\
2 & ECG5000 & 5000 & 140 & 5 \\
3 & Crop & 19412  & 46 &  24\\
4 & ElectricDevices & 16160 & 96 & 7 \\
5 & FreezerSmallTrain & 2878 & 301 & 2 \\
6 & HandOutlines & 1370 & 2709 & 2 \\
7 & InsectWingbeatSound & 2200 & 256 & 11 \\
8 & Mallat & 2400 & 1024 & 8 \\
9 & MixedShapesRegularTrain & 2925 & 1024 & 5 \\
10 & MixedShapesSmallTrain & 2525 & 1024 & 5 \\
11 & NonInvasiveFetalECGThorax1 & 3765 & 750 & 42 \\
12 & NonInvasiveFetalECGThorax2 & 3765 & 750 & 42 \\
13 & ShapesAll & 1200 & 512 & 60 \\
14 & SonyAIBORobotSurface2 & 980 & 65 & 2 \\
15 & StarLightCurves & 9236 & 84 & 2 \\
16 & UWaveGestureLibraryAll & 4478 & 945 & 8 \\
17 & UWaveGestureLibraryX & 4478 & 315 & 8 \\
18 & UWaveGestureLibraryY & 4478 & 315 & 8 \\
\end{tabular}
  \caption{Summary of UCR datasets used in the experiments. $n$ is the number of objects, and $L$ is the length or size of the object.}
  \label{tab:datasets}
\end{table}

\myparagraph{Datasets}
For evaluation, we use 18 datasets from the UCR Time Series Classification Archive~\cite{UCRArchive2018}. 
The IDs and sizes of the datasets are given in \cref{tab:datasets}. The dataset sizes are not large: this is because our algorithms require a correlation matrix as input, which uses $\Theta(n^2)$ memory. This high memory requirement significantly constrains the size of our datasets.
The correlation matrix for each dataset was computed by taking the Pearson correlation of each pair of time series. 

\begin{figure}[!t]
    \centering
    \includegraphics[width = 0.85\columnwidth]{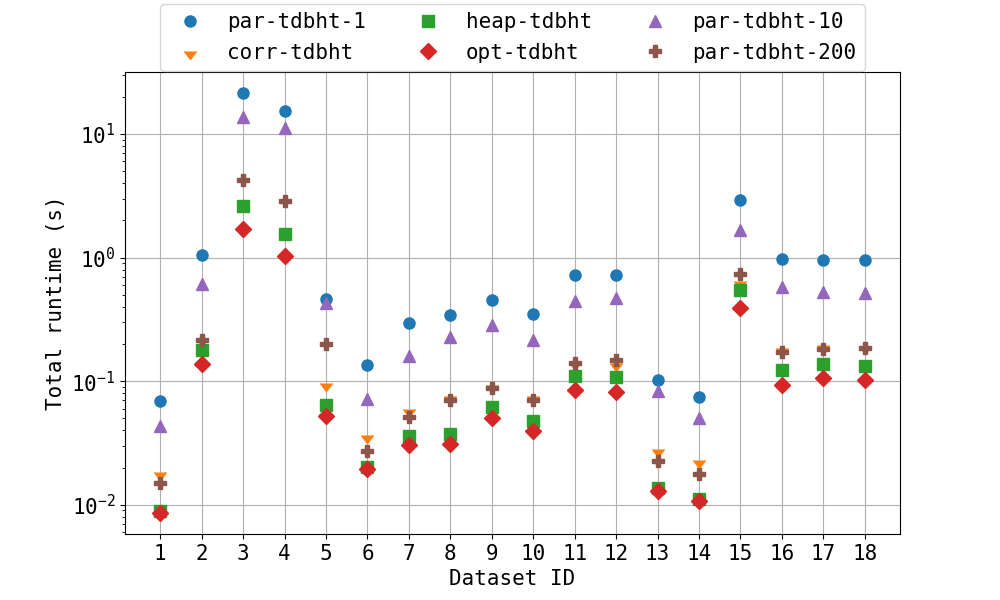}
    \caption{Parallel runtime of TMFG-DBHT methods on different datasets. }
    \label{fig:time}
\end{figure}

\subsection{Runtime}
 \cref{fig:time} shows the parallel running time of different TMFG-DBHT methods on our datasets.

First, we see that \textsc{par-tdbht-10} is faster than \textsc{par-tdbht-1} by a factor of 1.1--1.8.
We also see that \textsc{corr-tdbht} gives a 7--16x speedup over \textsc{par-tdbht-10}. 
Additionally, \textsc{heap-tdbht} gives a 1.1--1.36x speedup over \textsc{corr-tdbht} for the largest datasets, but it gives over 3x speedup for some of the smaller datasets. Finally, \textsc{opt-tdbht} gives a 1.42--1.52x speedup over \textsc{heap-tdbht} for the 3 largest datasets, and 1.03--1.3x speedup for the other datasets.
Compared to the previous state-of-the-art \textsc{par-tdbht-10}~\cite{yu2023parallel}, our fastest algorithm
\textsc{opt-tdbht} is 5.9x faster on average across all datasets, and is 5--10x faster on average for the three largest datasets, Crop, ElectricDevices, and StarLightCurves.
The reason why \textsc{opt-tdbht} has a larger speedup over \textsc{par-tdbht-10} on the largest datasets is due to the sorting step taking a larger fraction of the total time on the largest datasets, which we discuss below, 
and our algorithms are designed to reduce the parallel overhead of sorting.

\begin{figure}[!t]
    \centering
    \includegraphics[width = 0.5\columnwidth]{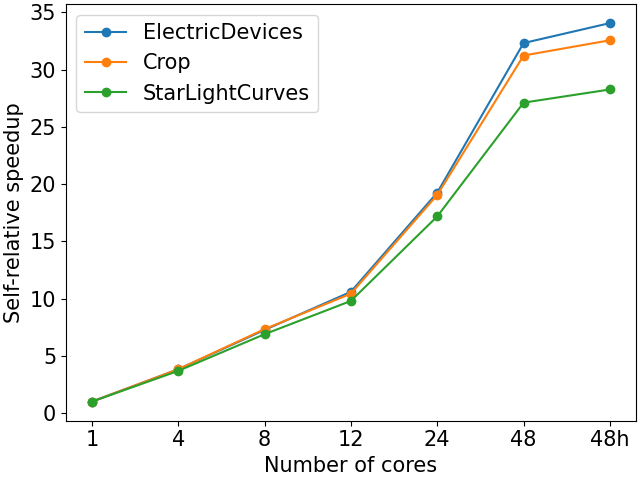}
    \caption{Self-relative parallel speedup of \textsc{opt-tdbht} on the three largest datasets for different numbers of cores. "48h" means 48 cores with 2-way hyper-threading. 
    }
    \label{fig:speedup}
\end{figure}

\begin{figure}[!t]
    \centering
    \includegraphics[width=0.5\columnwidth]{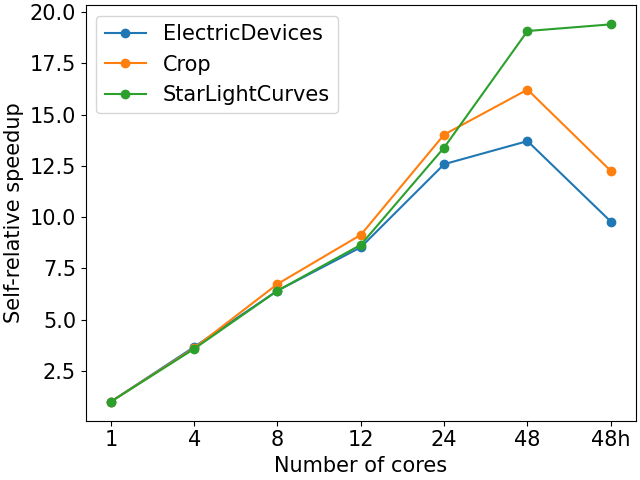}
    \caption{Self-relative parallel speedup of \textsc{par-tdbht-10} on the three largest datasets for different numbers of cores. "48h" means 48 cores with 2-way hyper-threading.}
    \label{fig:speedup-orig}
\end{figure}

We found that the self-relative parallel speedup of our fastest algorithm \textsc{opt-tdbht} on 48 cores was 7--34x across all datasets.
We show in \cref{fig:speedup} the self-relative speedup of \textsc{opt-tdbht} on the three largest datasets, ElectricDevices, Crop, and StarLightCurves. The algorithm is 27--33x faster when using 48 cores without hyper-threading compared to only using 1 core. By contrast, as shown in \cref{fig:speedup-orig}, \textsc{par-tdbht-10} is only 14--19x faster on these three datasets when using 48 cores.
Using 48 cores with 2-way hyper-threading did not give a dramatic speedup,
and sometimes led to a slowdown.


\begin{figure*}
\begin{subfigure}[b]{0.45\textwidth}
        \centering
        \includegraphics[width=0.95\linewidth]{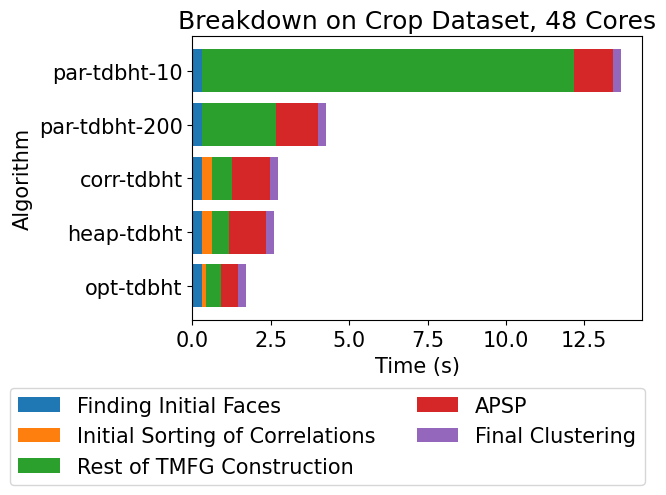}
    \end{subfigure}
~
\begin{subfigure}[b]{0.45\textwidth}
    \centering
    \includegraphics[width=0.95\linewidth]{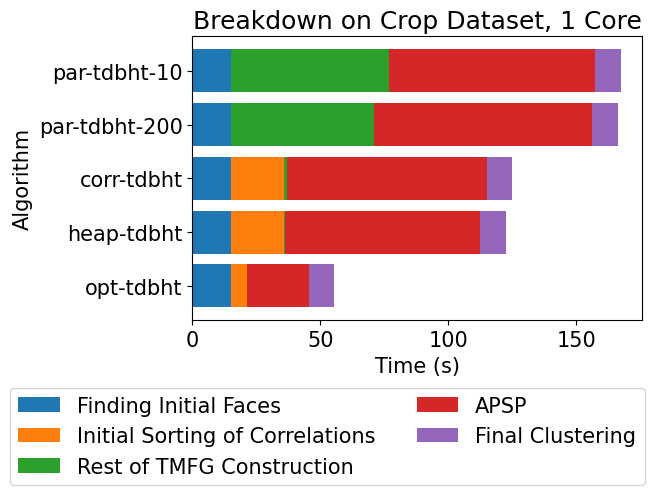}
\end{subfigure}
        \caption{Time breakdown of different algorithms on the Crop dataset, running on 48 cores with hyper-threading (left) and one core (right).}    \label{fig:breakdown}
\end{figure*}

\cref{fig:breakdown} (left) breaks down the various algorithms' parallel runtimes by step on the Crop dataset. The first step, finding initial faces, is part of TMFG construction: the only computationally intensive task it does is finding the best initial 4-clique. The initial sorting of correlations step sorts the rows of the input matrix in our new algorithms. 
The rest of the TMFG construction involves adding all other vertices to the TMFG. The APSP step computes all-pairs shortest paths on the TMFG, and the DBHT step finds the final clusters. On 48 cores with hyper-threading, the original \textsc{par-tdbht} algorithms' runtimes are clearly dominated by the vertex-adding portion of TMFG construction on Crop. \textsc{corr-tdbht} and \textsc{heap-tdbht} complete this step much faster, although these two algorithms need to sort the rows of the input correlation matrix. \textsc{opt-tdbht} has a faster sorting step than \textsc{corr-tdbht} and \textsc{heap-tdbht}, which can be attributed to the use of Google Highway, and a faster APSP step because it uses an approximate APSP algorithm.

In \textsc{par-tdbht-10}, the sorting steps of TMFG construction occur after finding initial faces. The runtime of post-initialization TMFG construction is dominated by this sorting step, as evidenced by the fact that the post-initialization TMFG construction step in \textsc{corr-tdbht} is much faster than the corresponding step in \textsc{par-tdbht-10}. 
 
On 48 cores, sorting during TMFG takes up roughly 87\% of the runtime of \textsc{par-tdbht-10} (including DBHT). In contrast, the initial sorting step in \textsc{corr-tdbht} takes 12\% of the total runtime.

\cref{fig:breakdown} (right) breaks down the runtimes by step on the Crop dataset when running on one core. 
On one core, sorting during TMFG construction takes up roughly 37\% of the entire runtime of \textsc{par-tdbht-10}, including DBHT, and the initial sorting step in \textsc{corr-tdbht} takes 13\% of the runtime.
There are no differences in runtime between the other steps of \textsc{par-tdbht-10} and the other steps of \textsc{corr-tdbht}.

Therefore, since there is a larger reduction in sorting time between \textsc{par-tdbht-10} and \textsc{corr-tdbht} when running on 48 cores than when running on one core, this implies that the sorting step in \textsc{corr-tdbht} reduces the overhead of sorting compared to the sorting step in \textsc{par-tdbht-10}. 

We now discuss the performance of TMFG specifically, including all sorting and initialization steps.
We found that TMFG construction time in \textsc{corr-tdbht} is 2--11x faster than TMFG construction time in \textsc{par-tdbht-10}, and TMFG construction time in \textsc{heap-tdbht} is 5--15x faster than TMFG construction time in \textsc{par-tdbht-10}, including for the three largest datasets. Specifically, TMFG construction in \textsc{heap-tdbht} is 10.5x faster on Crop, 15x faster on ElectricDevices, and 5.2x faster on StarLightCurves. TMFG construction in \textsc{heap-tdbht} is also 1.6--2.7x faster than \textsc{par-tdbht-200} on these three datasets (note that \textsc{par-tdbht-200} does not get good clustering quality, as we show later). 

We found that using manual vectorization, when applied on top of \textsc{heap-tdbht}, only produced a 0.97--1.07x speedup in TMFG construction for all datasets, including a 1.01--1.07x speedup in TMFG construction for the three largest datasets. 
Sorting with Google Highway on top of \textsc{heap-tdbht} and manual vectorization gives the \textsc{opt-tdbht} implementation, and leads to a 1.00-1.23x speedup in TMFG construction time for all datasets and a 1.18--1.23x speedup for the largest 3 datasets. Overall, the TMFG construction time in \textsc{opt-tdbht} is 6--20 times faster than the TMFG construction time in \textsc{par-tdbht-10}.

Additionally, we observe that switching from exact to approximate APSP speeds up the APSP step by a factor of 2--3 on most datasets, except for the smallest ones.

\subsection{Clustering Quality}
\begin{figure}[!t]
    \centering
    \includegraphics[width = 0.85\columnwidth]{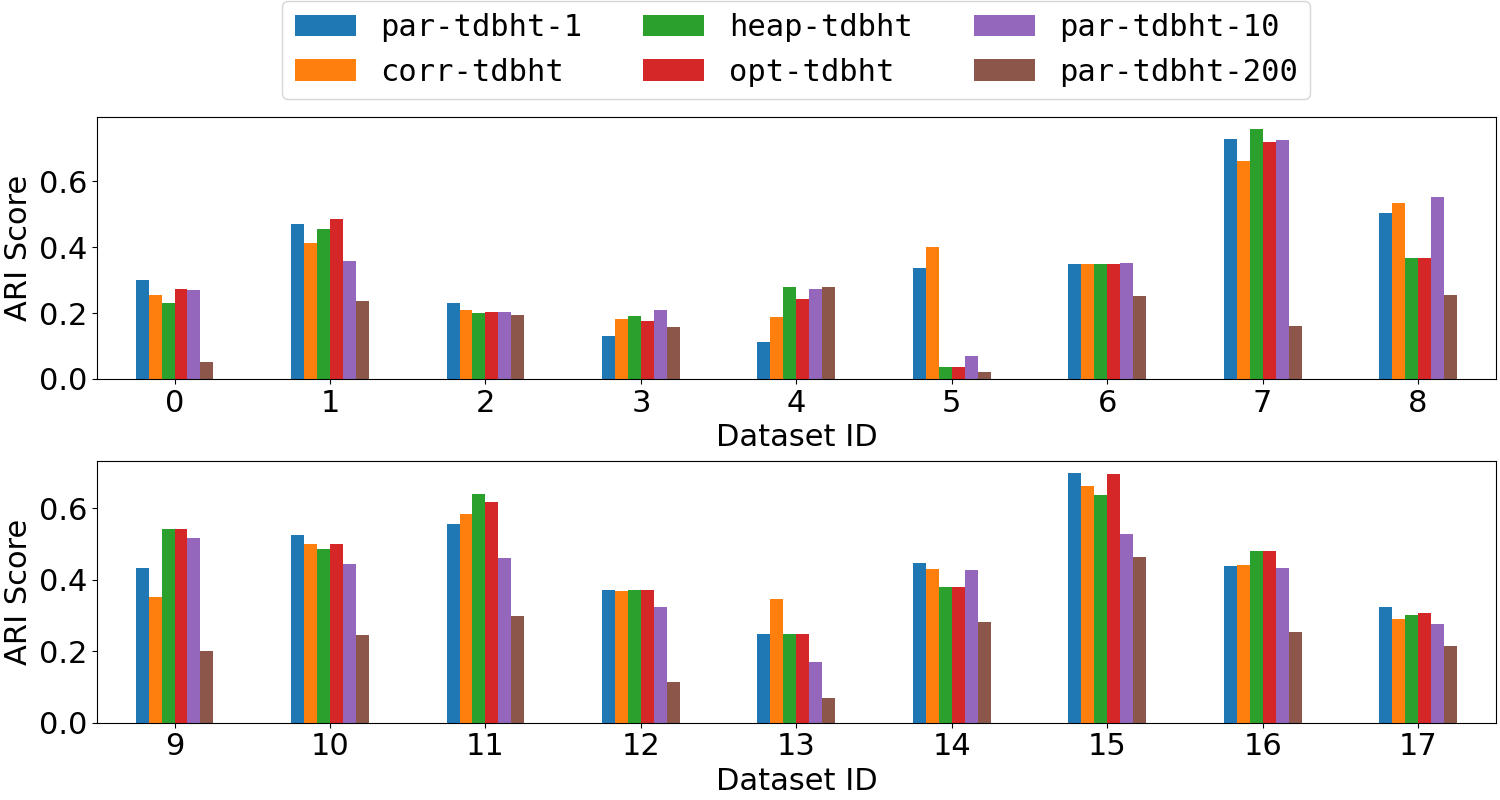}
    \caption{ARI scores of TMFG-DBHT methods. }
    \label{fig:score}
\end{figure}

We present the ARI scores of the different implementations in \cref{fig:score}.
We found that \textsc{opt-tdbht} only slightly changes the average ARI score compared to \textsc{par-tdbht} with prefix 10, with no significant dropoff.

For individual datasets, the ARI scores can vary widely between different methods. However, the average ARI scores for each method across all 18 datasets are similar to each other.
The average ARI score for \textsc{opt-tdbht} is 0.388. This is 6\% higher than the average ARI score for \textsc{par-tdbht-10}, which is 0.366. The average ARI score for \textsc{par-tdbht-200} is much lower, at 0.208. This is because a large prefix size causes the algorithm to create a graph with relatively low edge sums. The average ARI score for \textsc{par-tdbht-1} is slightly higher, at 0.400.


\begin{figure}[!t]
    \centering
    \includegraphics[width = 0.85\columnwidth]{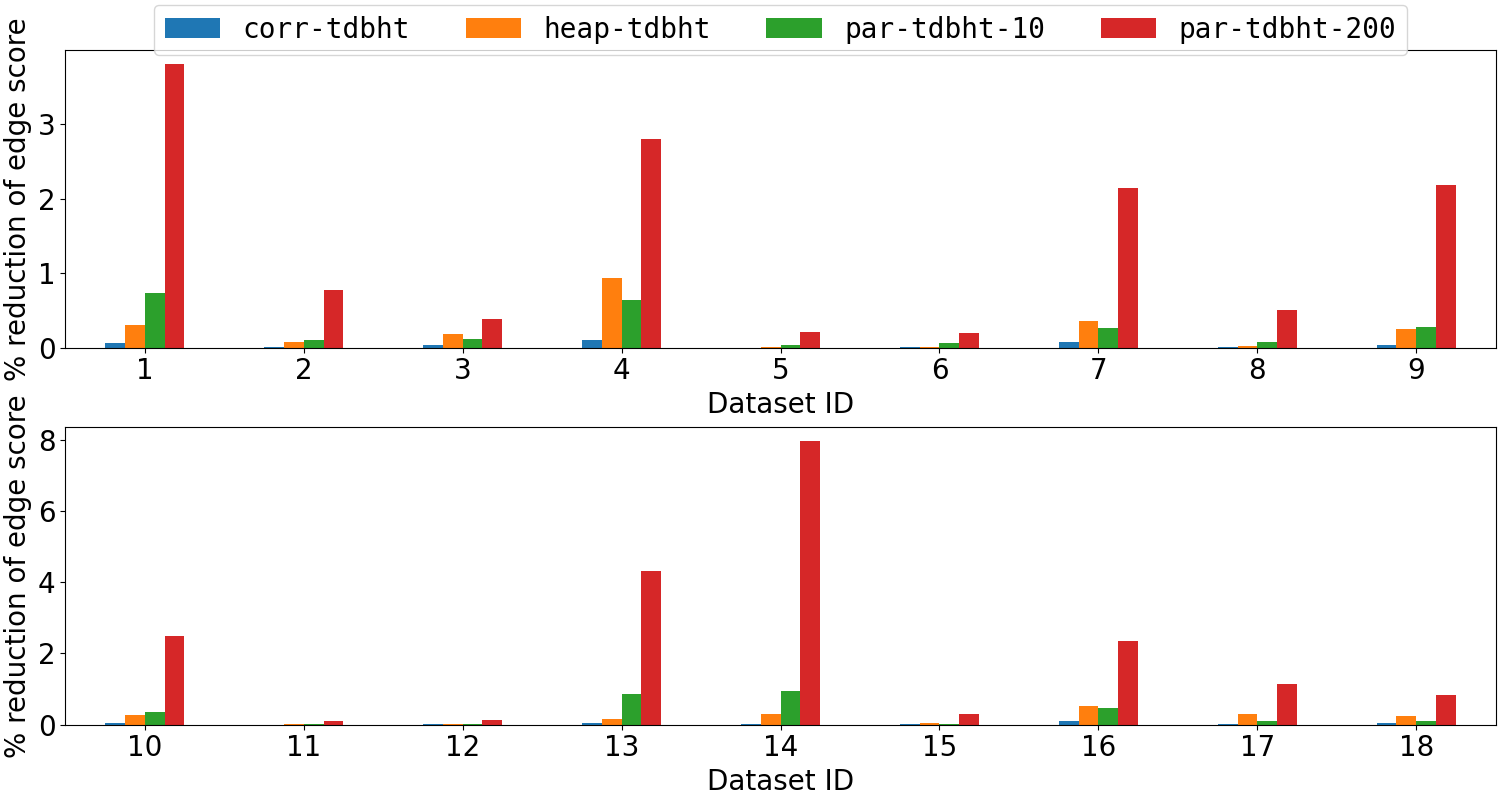}
    \caption{Percent reduction in edge sums for various methods compared to \textsc{par-tdbht-1}.}

    \label{fig:edge}
\end{figure}

\Cref{fig:edge} shows the TMFG edge sums of different methods compared to \textsc{par-tdbht-1}. 
The edge sums of the parallel methods are generally lower than that of \textsc{par-tdbht-1}, but 
\textsc{corr-tdbht} and \textsc{heap-tdbht} (which are equal to the scores for \textsc{opt-tdbht}, since the graph construction algorithm is the same for both) are comparable to \textsc{par-tdbht-10}.
In particular, the edge sum scores for \textsc{heap-tdbht} were all less than 1\% lower than those for \textsc{par-tdbht-1} and were between 0.4\% higher and 0.3\% lower than those of \textsc{par-tdbht-10}.


\section{Conclusion}
We presented an improved parallel algorithm and optimizations for TMFG+DBHT 
hierarchical clustering. We showed that our new algorithm achieves comparable accuracy to the state-of-the-art parallel algorithm, while achieving significant runtime speedups and good parallel scalability.
For future work, we are interested in studying the robustness of our algorithm and also making our algorithm dynamic.

\section{Acknowledgements}
This research was supported by
DOE Early Career Award \#DE-SC0018947,
NSF CAREER Award \#CCF-1845763, NSF Award \#CCF-2316235, Google Faculty Research Award, Google Research Scholar Award, and FinTech@CSAIL Initiative.

\enlargethispage{\baselineskip}

\bibliography{ref}
\end{document}